\documentstyle[12pt]{article}
\newcommand{\iiint}{{\int \! d^3 \! x \;}}

\renewcommand{\d}{\delta}
\newcommand{\g}{\gamma}
\newcommand{\e}{\varepsilon}

\newcommand{\m}{\mu}
\newcommand{\n}{\nu}

\renewcommand{\S}{\Sigma}

\newcommand{\r}{\rho}

\newcommand{\real}{{{\rm I} \kern -.19em {\rm R}}}
\newcommand{\tr}{{\rm {Tr} \,}}

\newcommand{\pa}{\partial}

\newcommand{\ie}{{{\em i.e.},\ }}

\newcommand{\equ}[1]{(\ref{#1})}
\newcommand{\SS}{{\cal S}}

\newcommand{\eq}{\begin{equation}}
\newcommand{\ee}{\end{equation}}
\newcommand{\eqn}[1]{\label{#1}\end{equation}}
\newcommand{\eea}{\end{eqnarray}}
\newcommand{\eqa}{\begin{eqnarray}}
\newcommand{\eqan}[1]{\label{#1}\end{eqnarray}}
\newcommand{\ba}{\begin{array}}
\newcommand{\ea}{\end{array}}
\newcommand{\eqac}{\begin{equation}\begin{array}{rcl}}
\newcommand{\eqacn}[1]{\end{array}\label{#1}\end{equation}}

\newcommand{\fud}[2]{{\displaystyle{\frac{\delta #1}{\delta #2}}}}

\newcommand{\lp}{\left(}
\newcommand{\rp}{\right)}
\newcommand{\lc}{\left[}
\newcommand{\rc}{\right]}

\newcommand{\f}{\phi}

\newcommand{\es}{\\[3mm]}

\newcommand{\journal}[4]{{\em #1~}#2\,(19#3)\,#4;}
\newcommand{\hpa}{\journal {Helv. Phys. Acta}}

\newcommand{\cmp}{\journal {Comm. Math. Phys.}}

\newcommand{\np}{\journal {Nucl. Phys.}}
\newcommand{\pl}{\journal {Phys. Lett.}}
\newcommand{\prep}{\journal {Phys. Reports}}

\newcommand{\nc}{\journal {Nuovo Cim.}}
\makeatletter
\@addtoreset{equation}{section}
\makeatother

\advance\hoffset by -0.6cm
\begin{document}
\advance\voffset by -2cm
\centerline{ {\LARGE {\sc
            {\bf UNIVERSIT\'E DE GEN\`EVE}}}}
\centerline{ \raisebox{0mm}{{\small SCHOLA GEVENENSIS MDLIX}} }
\vspace{72mm}

\centerline{\LARGE A short comment on }  \vspace{2mm}

\centerline{\LARGE the supersymmetric structure of }  \vspace{2mm}

\centerline{\LARGE Chern-Simons theory in the axial gauge}
\vspace{9mm}

\centerline{A. Brandhuber,  M. Langer and  M. Schweda  }
\centerline{{\small  Institut f\"ur Theoretische Physik} }
\centerline{{\small  Technische Universit\"at Wien }}
\centerline{{\small  Wiedner Hauptstra\ss e 8-10}}
\centerline{{\small  A-1040 Wien (Austria)}}
\vspace{4mm}

\centerline{O. Piguet$^1$\footnotetext[1]{Supported in part
                          by the Swiss National Science Foundation.}
 and  S.P. Sorella$^{1,2}$\footnotetext[2]{Supported in part
by the ''Fonds zur F\"orderung der Wissenschaftlichen Forschung'',
M008-Lise Meitner Fellowship.}  }
\centerline{{\small D\'epartement de Physique Th\'eorique}}
\centerline{{\small 24, quai Ernest Ansermet}}
\centerline{{\small CH -- 1211 Gen\`eve 4 (Switzerland)}}
\vspace{17mm}

\centerline{{\normalsize {\bf UGVA---DPT 1992/11--793}} }
\thispagestyle{empty}
\vfill
\pagebreak
\advance\voffset by 2cm
\null
\vspace{6cm}

\centerline{\Large{\bf Abstract}}\vspace{10mm}

\noindent The topological supersymmetry of the pure Chern-Simons
model in three dimensions is established in the case where
the theory  is defined in the axial gauge.
\setcounter{page}{0}
\thispagestyle{empty}

\vfill\pagebreak
\null
\section{Introduction}
One of the characteristic features of the Chern-Simons
theory~\cite{schwarz,witten},
and of more general topological field theories~\cite{bbrt},
is the supersymmetric structure \cite{brt,dgs,dlps,lps}
they possess, in certain gauges, and which is at the origin of
their ultraviolet finiteness\cite{dlps,lps}.
This supersymmetry is generated by the BRS operator -- and sometimes
also by the anti-BRS operator --
and by one -- or two -- operators carrying a Lorentz index.
For the three-dimensional
Chern-Simons theory such a supersymmetry
was known to hold in the Landau gauge~\cite{brt,dgs}. The
purpose of this note is to show that it also holds in the
homogeneous axial gauge.

The axial gauge is particular interesting for
the Chern-Simons theory~\cite{froehlich,martin,emery,kapustin},
and presumably for other topological theories,
since it is in this gauge
that the ultraviolet finiteness is the most
obvious due to the
complete absence of radiative corrections.
The choice of the axial gauge is also relevant for the study of
topological theories in manifolds with boundaries~\cite{moore,emery}.
Although the role of supersymmetry in such a
geometrical setup still needs a clarification, it is important to
establish its presence.

In this paper we treat only the gauge fixed classical theory.
However, due to the
aforementionned absence of radiative corrections in the axial gauge, our
results hold in fact for the full quantum theory.
\section{BRS invariance and supersymmetry}\label{section2}
The Chern-Simons gauge invariant action in three dimensions is given,
in the notation of~\cite{dlps}, by
\eq
\Sigma_{\rm CS}(A)=-{1\over 2}\iiint\epsilon^{\mu\nu\rho}
\,\tr\left( A_\mu\partial_\nu
   A_\rho +{2\over 3}\, g\,A_\mu A_\nu A_\rho \right)
\; .
\eqn{cs-action}
In the homogeneous axial gauge characterized by the constant
three-vector $n^\m$, one has to add the gauge fixing and  Faddeev-Popov
terms~\cite{martin,emery,leib-martin,kapustin}
\eq
\Sigma_{\rm gf}(A,d,b,c)=\iiint Tr(d\, n^\mu A_\mu
+b\, n^\mu D_\mu c) \;,
\eqn{gf-action}
where $d$ is the Lagrange multiplier field for the axial gauge
condition $n^\m A_\m=0$, $c$ and $b$ are
respectively the ghost and antighost fields
and $D_\m\cdot = \pa_\m\cdot + g[A_\m,\cdot]$ is the covariant
derivative.
The gauge group $G$ is assumed to be simple, and all the fields are Lie
algebra valued and belong to the adjoint representation:
\[
\f = \f_a \tau^a\ ,\quad{\rm for}\quad \f= A_\m,\ d,\ b,\ c \ ,
\]
where the $\tau^a$ are the generators of $G$, normalized in such a way that
\[
[\tau^a,\tau^b] = f^{abc}\tau^c\ ,\quad \tr(\tau^a\tau^b) = \d^{ab}\ .
\]
By construction the gauge fixed action
\eq
\S(A,d,b,c) = \S_{\rm CS} + \S_{\rm gf}
\eqn{tot-action}
is invariant under the nilpotent BRS transformations
\eq\ba{ll}
 sA_\mu &=-D_\mu c \;,\ \ \ \ \ \ \   sb=d \;,\es
             sc &=g\,c^2 \;,        \ \ \ \ \ \ \ \ \ \
  sd=0 \ ,
\ea\eqn{brs}

It is well known that the Chern-Simons action in the covariant Landau
gauge, besides the BRS symmetry, has other invariances generated by
anticommuting operators, which are
Lorentz vectors~\cite{brt,dgs,dlps}.
In the Landau gauge this vector symmetry has the remarkable property of
giving rise, together with the BRS operator, to a superalgebra
which, closing on the space-time translations, allows for a
supersymmetric interpretation of the model~\cite{dgs}.
Indeed, as shown by \cite{dlps}, this supersymmetry implies the
existence of a supercurrent which has been the starting point for the proof
of its perturbative finiteness.

It is remarkable that the
same supersymmetric structure is also present in the
homogeneous axial gauge considered here. It is indeed
straightforward to check that the action \equ{tot-action} is
invariant under the following infinitesimal transformations:
\eq\ba{ll}
v_\mu A_\nu &=\epsilon_{\mu\nu\rho}n^\rho b \;,
                     \ \ \ \ \ \ \   v_\mu b=0 \;,\es
            v_\mu c&=-A_\mu\;,     \ \ \ \ \
 \ \ \ \ \ \   v_\mu d=\partial_\mu b \;.
\ea\eqn{v-mu}
The only difference with the corresponding transformation laws in the
Landau gauge~\cite{dgs,dlps}
is that the partial derivative $\pa^\rho$ in the transformation of
$A_\m$ has been replaced by the gauge vector $n^\rho$.

As announced, these transformations,
the BRS transformations \equ{brs} and the translations
$\pa_\m$ obey the supersymmetry algebra
\eq
s^2=0 \;,\ \ \ \{ v_\mu ,v_\nu \} =0 \;,\ \ \ \{s,v_\mu \}=
\partial_\mu +({\rm field\ equations}) \ .
\eqn{alg-v-s}

It must be noted here that the above vector supertransformations are
different from the ones proposed in~\cite{martin}:
\eq\ba{ll}
\bar v_\mu A_\nu &=\epsilon_{\mu\nu\rho}n^\rho c \;,
                     \ \ \ \ \ \ \   \bar v_\mu b=-A_\mu \;,\es
            \bar v_\mu c&=0\;,     \ \ \ \ \
 \ \ \ \ \ \   \bar v_\mu d=\partial_\mu c \;,
\ea\eqn{v-bar-mu}
and which can be obtained from \equ{v-mu} just by interchanging the
ghost fields: $b \leftrightarrow c$. That the transformations
\equ{v-bar-mu} leave the action invariant follows from the fact that the
interchange of $b$ and $c$ is a (discrete) symmetry of the
theory. For
the same reason\cite{rheban}, to the BRS invariance \equ{brs} corresponds an
''anti-BRS'' invariance $\bar s$:
\eq\ba{ll}
 \bar sA_\mu &=-D_\mu b \;,\ \ \ \ \ \ \   \bar sb= g\,b^2\;,\es
             \bar s c &= d \;,        \ \ \ \ \ \ \ \ \ \
  \bar s d=0 \;.
\ea\eqn{brs-bar}
However, whereas in the Landau gauge
the corresponding operators $s$, $\bar s$, $v_\m$, $\bar v_\m$
and the translation operator $\pa_\m$ form
an $N=2$ supersymmetry algebra~\cite{dgs},
in the present case they don't give rise to a closed
algebra. For that
reason we focus only on the smaller $N=1$ algebra \equ{alg-v-s} of
the operators $s$, $v_\m$ and $\pa_\m$.

We remark that both supersymmetries $v_\m$ and $\bar v_\m$ originate
from the field equations for the components of the gauge
field which are tranverse with respect to
the gauge vector $n^\m$. These equations are obtained
from the field equation for $A_\r$, contracted with the $\e_{\m\n\r}$
tensor and the gauge vector $n^\n$. They
take the form of a linearly broken Ward identity, after use of the
gauge condition
\eq
\fud{}{d} \S = n^\m A_\m\ ,
\eqn{gauge-cond}
where $\S$ is   the gauge fixed action \equ{tot-action}.
They read
\eq
\lp \epsilon_{\m\n\r} n^\n \fud{}{A_\r}
  + D_\m\fud{}{d} \rp \S = n^\n\pa_\n A_\m\ .
\eqn{susy-loc}
The invariance of the action under the two supersymmetries \equ{v-mu}
and\equ{v-bar-mu} is easily recovered by looking on the functional identities
obtained by multiplying
\equ{susy-loc} either by the antighost field $b$ or by the ghost field
$c$, and then by integrating on space-time.
\section{Functional identities and off-shell algebra}
BRS invariance as well as supersymmetry may be expressed by functional
identities upon the introduction of the external fields $\g^\m$ and $\tau$
coupled to the nonlinear BRS transformations of $A_\m$ and $c$,
respectively, \ie upon the addition, to the action \equ{tot-action}, of
the terms
\eq
\S_{\rm ext} = \tr \iiint \left( -\gamma^\mu D_\mu c
                                              +g\,\tau c^2\right)\ .
\eqn{ext-action}
Thus the complete classical action
\eq
\S(A,d,b,c,\g,\tau) =  \S_{\rm CS} + \S_{\rm gf}  + \S_{\rm ext}\ ,
\eqn{complete-action}
obeys the Slavnov identity
\eq
{\cal S}(\Sigma )=\tr \iiint \left( {\delta\Sigma
\over\delta A_\mu}{\delta\Sigma\over\delta\gamma^\mu}
             +{\delta\Sigma\over\delta c}
{\delta\Sigma\over\delta\tau}
             +d {\delta\Sigma\over\delta b}\right) =0\ .
\eqn{slavnov}
for BRS invariance, and the broken Ward identity
\eq
{\cal V}_\mu \S = \Delta_\mu^{\rm cl}
\equiv \iiint \tr(-\gamma^\nu\partial_\mu A_\nu
+\epsilon_{\mu\nu\rho}\gamma^\nu
 n^\rho d+\tau\partial_\mu c)\ ,
\eqn{susy-wi}
where
\eq
{\cal V}_\mu
\equiv \tr\iiint \left( \epsilon_{\nu\mu\rho}\left( \gamma^\rho
-n^\rho b\right) {\delta\over\delta A_\nu}
                -A_\mu{\delta\over\delta c}+\partial_\mu
b{\delta\over\delta d}
                -\tau{\delta\over\delta\gamma^\mu}\right)\ ,
\eqn{susy-wi-op}
for the supersymmetry \equ{v-mu}.
The term $\Delta_\mu^{\rm cl}$ in the right-hand-side of
\equ{susy-wi}, being linear in the quantum fields, represents a classical
breaking which will be left unchanged by the renormalization. This
feature is commun to a large class of topological
theories~\cite{dlps,ms,lps}.

Again supersymmetry, expressed by the broken
Ward identity \equ{susy-wi}, follows from the
local Ward identity \equ{susy-loc} which, in the presence of the
external fields, takes the form
\eq
{\cal D}_\m (x) \S = n^\n\pa_\n A_\m
             + g \e_{\m\n\r} n^\n \{\g^\r, c\}\ ,
\eqn{susy-loc-s}
with
\eq
{\cal D}_\m(x) \equiv \epsilon_{\m\n\r} n^\n \fud{}{A_\r}
  + D_\m\fud{}{d}    \ .
\eqn{op-d_mu}
In order to see this, one  has to multiply \equ{susy-loc-s} by $b$,
integrate over space-time, and use the gauge condition
\eq
\fud{\S}{d} = n^\m A_\m\ ,
\eqn{gauge-cond-s}
as well as the equation
\eq
\fud{\S}{c} + g\lc b,\fud{\S}{d} \rc = n^\n\pa_\n  b  - \pa_\m\g^\m
   - g\lc A_\m,\g^\m \rc + g\lc c,\tau \rc\ .
\eqn{}
The latter identity is a local form, valid in the axial gauge, of the
ghost equation of~\cite{bps}.

The on-shell algebra given by the anticommutators \equ{alg-v-s},
together with BRS nilpotency, is promoted to an off-shell nonlinear
algebra. Given an arbitrary functional
$F$ of the fields $A_\mu$, $d$, $b$, $c$, $\gamma^\mu$ and $\tau$,
this algebra reads
\eq\ba{l}
 {\cal S}_F {\cal S}(F )=0\;,\ \ \ \ \ \ \
 \ \ \{ {\cal V}_\mu ,{\cal V}_\nu\} =0 \;,\es
             {\cal V}_\mu {\cal S}(F)
+{\cal S}_F ({\cal V}_\mu F
- \Delta_\mu^{\rm cl})
            ={\cal P}_\mu F \;,
\ea\eqn{off-shell-alg}
where ${\cal P}_\mu $ is the translation Ward operator
\[
{\cal P}_\mu  =\iiint \sum_{\rm All \; fields}
 \partial_\mu \phi {\delta\over\delta \phi }   \;,
\]
and $\SS_F$ is the $F$-dependent linearized
Slavnov operator
\eq
{\cal S}_F=\tr \iiint \left( {\delta F
\over\delta A_\mu}{\delta\over\delta\gamma^\mu}
                  +{\delta F\over\delta
\gamma^\mu}{\delta\over\delta A_\mu}
                 +{\delta F\over\delta c}
{\delta\over\delta\tau}
                  +{\delta F\over\delta\tau}
{\delta\over\delta c}
               +d {\delta\over\delta b}\right)\  .
\eqn{slavnov-lin}

Moreover, if the functional $F$ is a solution of the Slavnov
identity \equ{slavnov}
and of the supersymmetry Ward identity \equ{susy-wi}, then the
linearized Slavnov operator and the supersymmetry Ward operator obey the
anticommutation rules
\eq
{\cal S}_F {\cal S}_F =0\;,\ \ \ \ \ \ \ \ \
\ \ \ \ \{ {\cal S}_F ,{\cal V}_\mu \}
={\cal P}_\mu \ .
\eqn{lin-alg}
\section{Conclusions}
We have shown that the Chern-Simons theory in the axial gauge is
not only invariant under the BRS transformations $s$
and a type of anti-BRS transformations $\bar s$:
it also possesses two supersymmetries $v_\mu$ and $\bar v_\mu$.
Moreover the two sets of operators $\{s,v_\mu\}$ and
$\{\bar s,\bar v_\mu\}$ generate two separate superalgebras of the
Wess-Zumino, $N=1$, type (see \equ{alg-v-s}). They are related by a
discrete transformation which consists in the interchange of the ghost
and of the antighost fields. However, contrarily to what happens in the
covariant Landau gauge, their union does not generate a $N=2$
supersymmetry algebra.

We have concentrated on the algebra generated by $\{s,v_\mu\}$, writing
the corresponding Ward identities and the off-shell algebra obeyed by the
functional generators. We have also found that the supersymmetry $v_\mu$
results from a local Ward identity (see \equ{susy-loc-s}) which is
peculiar to the axial gauge.


\begin{thebibliography}{99}
\bibitem{schwarz} A.S. Schwarz, Baku International Topological
                Conference, Abstracts, vol. 2, p. 345, (1987);
\bibitem{witten} E. Witten,
                  \cmp{121}{89}{351}
\bibitem{bbrt} D. Birmingham, M. Blau, M. Rakowski and G. Thompson,
        \prep{209}{91}{129}
\bibitem{brt} D. Birmingham, M. Rakowski and G. Thompson,
     \np{B329}{90}{83}
\bibitem{dgs} F. Delduc, F. Gieres and S.P. Sorella,
     \pl{B225}{89}{367}
\bibitem{dlps} F. Delduc, C. Lucchesi, O. Piguet and S.P. Sorella,
        \np{B346}{90}{313}
\bibitem{lps}  C. Lucchesi, O. Piguet and S.P. Sorella,
    {\em Renormalization and finiteness of topological BF theories},
   preprint MPI-Ph/92-57; UGVA-DPT 1992/07-773, to appear in
   {\em Nucl. Phys.} B;
\bibitem{froehlich} J. Fr\"ohlich and C. King, \cmp{126}{89}{167}
\bibitem{martin} C.P. Martin, \pl{B263}{91}{69}
\bibitem{emery} S. Emery and O. Piguet, \hpa{64}{91}{1256}
\bibitem{kapustin} A.N. Kapustin and P.I. Pronin, \pl{B274}{92}{363}
\bibitem{moore} G. Moore and N. Seiberg,
     \pl{B220}{89}{422}
\bibitem{leib-martin} G. Leibbrandt and C.P. Martin, \np{B377}{92}{593}
\bibitem{rheban} A. Rheban, \nc{100\,A}{88}{713}
\bibitem{ms}
        N. Maggiore and S.P. Sorella, \np{B377}{92}{236}\\
        N. Maggiore, PhD Thesis (1992);\\
        E. Guadagnini, N. Maggiore and S.P. Sorella,
        \pl{B255}{91}{65}\\
        N. Maggiore and S.P. Sorella, {\em Perturbation theory for
        antisymmetric tensor fields in four dimensions},
        preprint GEF-Th-6/1992; UGVA-DPT 1992/04-761,
        to be publ. in {\em Int. J. Mod. Phys.} A;
\bibitem{bps} A. Blasi, O. Piguet and S.P. Sorella, \np{B356}{91}{154}
\end{thebibliography}
\end{document}